\documentstyle[12pt,fleqn]{article}

\begin{document}
\date{}
%E-mail: vn@vg3025.spb.edu
\author{V.N. Gorbachev, A.I. Trubilko}
\title{Teleportation of entanglement for continuous variables}

\maketitle
\begin{abstract}
Teleportation of a  pure two particle entangled state
of continuous variables  by triplet of the 
Greenberger-Horne-Zeilinger form is considered. 
 The three-particle basis needed for a joint measurement is found.
It describes a measurement of momentum one of single particle and total moment and relative
position of the two others. Optical realization using squeezed state of the light is discussed.

\end{abstract}

\section{Introduction}

The main resources of quantum teleportation proposed by Bennett et al \cite{1} and demonstrated in experiments
\cite{2}, \cite{21} are an EPR pair and Bell measurement. If a sender or Alice wishes to teleport
a two qubit state   one EPR pair and two bits of classical information are required.
Because of the linearity of quantum mechanics,
teleportation of N particles can be achieved one-by-one using N EPR pairs, 
however if the N particle state is entangled it needs another resource.

Lee et al \cite {2} have considered a protocol of a two-spin pure state
in noisy channel modeled via two Werner states instead of two EPR pairs.
It has been shown that the one by one teleportation
of entanglement cannot be perfect even when the channel is quantum correlated.
Indeed in the case of multiqubit entanglement the one by one protocol 
can be modified \cite{4}. It results in
two entangled qubits can be  teleported by a triplet of the Grinberger-Horne Zeilinger (GHZ) form using 3/2
bits per particle by contrast 2 bits per particle if two EPR pairs are involved.
Karlsson et al \cite {5} have proposed GHZ to transmit single qubit for two receivers
however one receiver and only one of them  can recover an unknown state.

Teleportation of continuous variables proposed by Vaidmann  \cite{6} has been developed by Braunstein et al
\cite{7}, \cite{8} and many others.  This version is very attractive because of its
optical realization where properties of squeezed light are exploited to the full
as it has been demonstrated in the  optical experiment \cite{21} using the light of the optical parametric
oscillator as the source of EPR pair.

In this work we consider how to teleport two particle entanglement of 
continuous variables  by triplet GHZ. In this scheme Alice and two receivers Bob and Claire spatially separated share
an entangled state of the form of GHZ. Alice wishes to send an unknown two particle entangled state.
She performs a three particle joint measurement which outcomes she transmits to
both receivers as 3 bits of classical information.
We apply formalism presented in \cite{8} and find that the main feature of
the discrete version of teleportation is reproduced in continuous variables. So that first,
the required three particle basis for joint measurement
must be two particle entangled by contrast the maximally entangled basis in the
one by one protocol, second an unknown two particle state can be recovered
by two receivers together. We discuss optical realization based on squeezed light.

\section{Teleportation of continuous variables}

For main resources of teleportation such as ERP pair and the Bell basis in the continuous case
representation of the position and momentum basis eigenstates  $|x\rangle$, and
$|p\rangle$
is convenient. Using the complete and orthogonal set of
$|x\rangle$, any N - particle state can be written in the form
\begin{equation}
|A\rangle_{1,..N}=\int dx_{1}...dx_{N}A(x_{1},..x_{N})|x_{1}\rangle
...|x_{N}\rangle
\label{2}
\end{equation}
The continuous Bell states can be obtained from (\ref{2}),  if N=2 and $A(x_{1},x_{2})=\delta (x_{1}-x_{2}-Q)\exp(2iPx)/\sqrt{\pi}$
\begin{equation}
|\Psi(P,Q)\rangle=\frac{1}{\sqrt{\pi}}\int dx\exp (2iPx)|x\rangle
|x-Q\rangle
\label{4}
\end{equation}
The functions
$|\Psi(P,Q)\rangle$ are eigenstates of total momentum
$P_{P}=(p_{1}+p_{2})/\sqrt{2}$
and relative position $X_{Q}=(x_{1}-x_{2})/\sqrt{2}$ of two particles
for which eigenvalues are $P$ and
$Q$. The continuous Bell states are complete and orthogonal
\cite{11}
\begin{eqnarray}
\int dPdQ|\Psi(P,Q)\rangle \langle\Psi(P,Q)| =1
\nonumber
\end{eqnarray}
\begin{equation}
\langle \Psi(P,Q)|\Psi(P',Q')\rangle = \delta (P-P')\delta(Q-Q')
\label{5}
\end{equation}
For $Q=0$ and $P=0$ one can find a maximally entangled state
$|\Psi(0,0)\rangle$ or EPR pair.

Let Alice wishes to teleport an unknown state $|A\rangle_{1}$
of a particle 1 to Bob. It needs an EPR pair
$|\Psi(0,0)\rangle_{23} $  of particles 2, 3 shared with A and B and
a joint measurement of particles 1 and 2 in the Bell basis
$|\Psi(P,Q)\rangle_{12}$.
Writing the wave function of the combined system in the form
\begin{equation}
|A\rangle_{1}|\Psi(0,0)\rangle_{23}=\frac{1}{\pi}\int dPdQ|\Psi(P,Q)\rangle _{12}
|B\rangle_{3}
%U^{\dagger}(P,Q)|A\rangle_{3}
\label{6}
\end{equation}
one finds
the state of the Bob particle $|B\rangle$ to be connected with an
initial state $|A\rangle$ by unitary transformation \cite{8}
\begin{equation}
|B\rangle =U_{B}(P,Q)|A\rangle
\label{7}
\end{equation}
\begin{equation}
U_{B}=\int dx \exp(2iPx)|x\rangle \langle x-Q|
\label{71}
\end{equation}
The series expansion given by (\ref{6}) describes a joint measurement
of the total momentum $P_{P}$
and the relative position $X_{Q}$ of particle 1 and 2.
If outcome $P$, $Q$ is obtained then the Bob particle 3
is projected into state that
can be transformed unitary to an unknown teleported state  $|A\rangle$.

It is well known realization of the  presented continuous protocol in quantum optics where
electromagnetic field is described by the photon operators of creation
and annihilation $a^{\dagger}$ and $a$.
 The canonical momentum and position of the light are the quadrature operators
$X(\theta)=(a\exp(i\theta) +h.c.)/2$,  when $p=X(-\pi/2)$ and  $x=X(0)$. They can be measured
by the balance detection scheme where signal and a local field oscillator are mixed and the
difference of two photocurrents is measured. As the source of EPR pair
a subthreheld optical parametric oscillator (OPO) produced pairs
of correlated photons can be used.
For a simple model of OPO its output is
$a=a_{0}\cosh r+a^{\dagger}_{0}\sinh r$,  where  $a^{\dagger}_{0},a_{0}$, are input operators,
r is the squeezing parameter. Output state of the light can be squeezed  over one of the quadrature,
so that we find momentum squeezing if  $r>0$ and position squeezing if   $r<0$.
For perfect squeezing $|r|\to\infty$ the noise of the light described by the standard deviation
of $\langle (\Delta X)^{2}\rangle$, where  $\Delta X =X-\langle X\rangle$,
$\theta=0,-\pi/2$, can be suppressed
up to zero in a frequency range: $\langle (\Delta X)^{2}\rangle \to 0$.
Such property arises for instance when the averaging state is  eigenstate of
the considered operator $X$, then squeezing results in entanlement.

EPR pair of the state $|\Psi(0,0)\rangle_{23}$ can be prepared by splitting two
beams from two OPO like
$b_{j}=a_{j0}\cosh r_{j}+a^{\dagger}_{j0}\sinh r_{j}$, $j=2,3$,
where one of the beams has momentum squeezing  $r_{2}<0$, and the other is position squeezed
$r_{3}>0$.  Then in the output of the 50\% beamsplitter one can find superposition of the form
$a_{2}=(b_{2}+b_{3})/\sqrt{2}$,
$a_{3}=(b_{2}-b_{3})/\sqrt{2}$. For perfect squeezing
$|r_{2}|$, $r_{3}\to \infty$ the needed properties of EPR pair arise
\begin{equation}
x_{2}-x_{3}=0
\label{8}
\end{equation}
\begin{equation}
p_{2}+p_{3}=0
\end{equation}
where  $x_{2,3}$ and $p_{2,3}$ are position and momentum operators of particles
2 and 3.

In \cite{8} the Heisenberg picture is presented to describe teleportation based on
squeezed light. For Bob particle 3 the following operator identities are fair
\begin{equation}
x_{3}=x_{1}-(x_{2}-x_{3})-\sqrt{2}X_{Q}
\label{9}
\end{equation}
\begin{equation}
p_{3}=p_{1}+(p_{2}+p_{3})-\sqrt{2}P_{P}
\label{10}
\end{equation}
where  $X_{Q}=(x_{1}-x_{2})/\sqrt{2}$ and
$P_{P}=(p_{1}+p_{2})/\sqrt{2}$ are operators of joint measurement of particles
1 and 2 in the Bell basis $|\Psi(P,Q)\rangle$. If the  outcome  is $P$,  $Q$, in
(\ref{9}) and  (\ref{10}) it results in $X_{Q}\to Q$ and $P_{P}\to P$. Then taking into account
properties of EPR pair given by (\ref{8}) one find operators of particle 3 to be equal up to unitary transformation
to operators of teleported particle 1:
$x_{3}=x_{1}-\sqrt{2}Q$ and $p_{3}=p_{1}-\sqrt{2}P$.
Note here the squeezed light produces entanglement.

\section{Teleportation of entangled state}

>From linearity, teleportation of N particle can be achieved one by one using N EPR pairs,
however in particular case of entangled state it needs a less expensive 
channel.
Considering in (\ref{2}) two particles for which
$ A(x_{1},x_{2}) = A(x_{1})\delta (x_{1}-x_{2}-q)$
one can find an entangled state of the form
\begin{equation}
|A\rangle_{12}=\int dx A(x)|x\rangle |x-q\rangle\label{11}
\end{equation}
It is a eigenstate  of the relative positions  operator  $x_{1}-x_{2}$ with
eigenvalue q.

Two particle-entangled state (\ref{11}) can be teleported by the GHZ triplet
of particles 3,4 and 5 shared with Alice and two receivers Bob and Claire, where Alice has
particle 3 and the others 4,5 are in the Bob and Claire site. Let Alice performs
a join measurement of particles 1,2,3 which outcome is $K$.
It results that the state of particles 4 and 5 is entangled and projected into state
 $|BC\rangle $ that can be transformed  from the  initial state $|A\rangle$
by a map
\begin{equation}
|BC(K)\rangle =U_{BC}(K)|A\rangle 
\label{12}
\end{equation}
For the teleportation protocol operator $U_{BC}$ must be unitary. More of then let this operator
 be factorized when acting on the considered entangled state
\begin{equation}
U_{BC}|A\rangle =U_{B}U_{C}|A\rangle
\label{121}
\end{equation}
In other words we assume that only both receivers B and C can retrieved an unknown state but one of them cannot do it.

Operator  $U_{BC}$ is defined by the basis of the joint measurement.
It would be possible to expect that the task can be accomplished by the three-particle basis
of the form
\begin{equation}
|\Psi(P,Q,R)\rangle =\frac{1}{\sqrt{\pi}}
\int dx \exp(3ixP)|x\rangle |x-Q\rangle |x-R\rangle
\label{13}
\end{equation}
where three particles are entangled.
The introduced functions being eigenfunctions of the total momentum operator $p_{1}+p_ {2}+p_{3}$ and relative positions
as  $x_{1}-x_{2}$, $x_{1}-x_{3}$ with eigenvalues P and Q,R are complete and orthogonal set.
If $Q,R,P =0$ one can find the maximally entangled state of the GHZ form
\begin{equation}
|GHZ\rangle =\frac{1}{\sqrt{\pi}}
\int dx |xxx\rangle
\label{14}
\end{equation}
Using this basis and triplet GHZ to teleport an unknown entangled state
$|A\rangle$
it results in the operators $U_{B}$, $U_{C}$ of the form
\begin{eqnarray}
U_{B}(P,R)&=&\int dx\exp(3iPx)|x-R\rangle \langle x|=U^{\dagger}
\nonumber
\\
U_{C}(R,Q)&=&\int dx |x-R\rangle \langle x-Q| \delta(Q-q)
\nonumber
\end{eqnarray}
where only one of them is unitary. As for discrete case considered in
\cite{4} the maximally entangled basis does not result in the solution of the problem
for continuous variables.

To teleport an entanglement of the form   (\ref{11}) it needs a three - particle basis where
two particles are entangled only.
This condition can be satisfied by the set
$ \pi_{1(23)}=\{|b\rangle_{1} |\Psi(P,Q) \rangle_{23} \} $ where wave functions
$|b\rangle_{1}$  form a basis in the Hilbert space of the particle 1
\begin{equation}
\int db |b\rangle \langle b|=1
\label{20}
\end{equation}
and
$|\Psi(P,Q)\rangle_{23}$ is Bell state
of particle 2 and 3 given by (\ref{4}). With the use of
basis $\pi_{1(23)}$
one can write
the combined state as
\begin{equation}
|A\rangle_{12}|EPR\rangle_{345}=\frac{1}{\pi^{3}}
\int dbdPdQ|b\rangle_{1}|\Psi(P,Q)\rangle_{23}U_{B}(P,Q)
U_{C}(b,Q,q)|A\rangle_{45}
\label{21}
\end{equation}
where the unitary operator $U_{B}$ is defined  by (\ref{7}) and
\begin{equation}
U_{C}(b,Q,q)=\sqrt{\pi}\int dx \langle x+q|b\rangle |x\rangle\langle x-Q|
\label{22}
\end{equation}
Operator  $U_{C}$ is unitary if and only if  $\pi|\langle b|x+q\rangle|^{2}=1$.
It enables to establish the basis  (\ref{20}) obviously resulting
in a set of eigenfunctions of the momentum operator of the particle 1.
Now the wave functions from  $\pi_{1(23)}$ in (\ref{21}) take the form
$\{|p\rangle_{1}|\Psi(P,Q)\rangle_{23}\}$  being eigenfunctions of the momentum operator
of particle 1 $p_{1}$ and operators of particles 2 and 3 such
as  the relative position  $Y_{Q}=(x_{2}-x_{3})/\sqrt{2}$ and
total momentum  $\Pi_{P}=(p_{2}+p_{3})/\sqrt{2}$ with eigenvalues  $p$, $P$
and $Q$.
As it follows from (\ref{21})
for outcome of measurement to be $p$, $P$ and $Q$ an unknown state is connected to the
state of particles 4 and 5 by unitary transformation
\begin{equation}
|BC(p,P,Q,q)\rangle=U_{B}(P,Q)U_{C}(p,P,Q,q)|A\rangle
\label{23}
\end{equation}
Then after  acting by unitary operations  $U_{B}^{\dagger}$, and $U_{C}^{\dagger}$
both receiver B and C recover unknown state.

Consider how to prepare the two-particle entanglement in the form of (\ref{11}) that can be teleported with
the use of triplet GHZ.
If to mix by beamsplitter two states like $\int A(x)|x\rangle dx$ and
$|x=0\rangle$ one finds
the desirable state after beamsplitter.
As entanglement  $|A\rangle_{12}$ is eigenstate of operator
$X_{Q}$ or some difference of quadratures  then  their uncertainties are equal to zero in this state.
Such property has the two mode light with the relative position or a difference quadratures
to be squeezed.
In the Heisenberg picture as for EPR pair (\ref{8}) the position operators of modes satisfy equation
\begin{equation}
x_{1}-x_{2}=0
\label{24}
\end{equation}

In the same way the GHZ triplet can be prepared from squeezed light \cite{8}. It needs mixing of
the tree modes of light we denote as particles 3,4 ,5 by two nonabsorbing beamsplitters described by their matrix
\begin{eqnarray}
T=
\left(
\begin{array}{cc}\cos\theta&\sin\theta\\
\sin\theta&-\cos\theta
\end{array}
\right)
\label{25}
\end{eqnarray}

Let the modes 3 and 4  are input of the first beamsplitter for which  $\sin\theta=1/\sqrt{3}$ and
a beam of one of its output is mixed with mode 5 by the second 50\% beamsplitter.
Then the bosonic operators of input modes
$b^{\dagger}_{j}$, $b_{j}$, j=3, 4, 5  and output operators $a^{\dagger}_{j}$, $a_{j}$
are connected by unitary transformation
\begin{eqnarray}
a_{3}&=&\frac{1}{\sqrt{3}}b_{3}+\sqrt{\frac{2}{3}}b_{4}
\nonumber
\\
a_{4}&=&\frac{1}{\sqrt{3}}b_{3}-\frac{1}{\sqrt{6}}b_{4}
+\frac{1}{\sqrt{2}}b_{5}
\label{26}
\\
a_{5}&=&\frac{1}{\sqrt{3}}b_{3}-\frac{1}{\sqrt{6}}b_{4}
-\frac{1}{\sqrt{2}}b_{5}
\nonumber
\end{eqnarray}
Let input is  the light of OPO for which
$b_{j}=a_{j0}\cosh r_{j}+a^{\dagger}_{j0}\sinh r_{j}$, j=3, 4, 5,
where modes 4 and 5 have the position squeezing ($r_{4,5}>0$)
and mod 3 has the momentum squeezing ($r_{3}<0$). It follows from (\ref{26})
 that in the case of perfect squeezing $r_{4,5}$, $|r_{3}|\to\infty$ there is the triplet of
GHZ with properties
\begin{eqnarray}
x_{3}-x_{4}&=&0
\nonumber
\\
x_{3}-x_{5}&=&0
\label{27}
\\
p_{3}+p_{4}+p_{5}&=&0
\nonumber
\end{eqnarray}

Now we can  present the Heisenberg picture of teleportation entanglement (\ref{11})
by triplet GHZ. Series expansion  of the form (\ref{21})
describes a measurement of the observable such as
momentum $p_{1}$, relative position $Y_{Q}=(x_{2}-x_{3})/\sqrt{2}$ and total momentum $\Pi_{P}=(p_{2}+p_{3})/\sqrt{2}$,
where $p_{j}$ and  $x_{j}$ are the momentum and position operators of modes $j=1,2,3$.
Similarly  to (\ref{9}) è  (\ref{10}) we can write identities for modes 4 and 5
\begin{eqnarray}
x_{4}&=&x_{2}+(x_{4}-x_{3})-\sqrt{2}Y_{Q}
\nonumber
\\
x_{5}&=&x_{1}+(x_{2}-x_{1})+(x_{5}-x_{3})-\sqrt{2}Y_{Q}\label{28}
\end{eqnarray}
\begin{eqnarray}
p_{5}&=&p_{2}+(p_{3}+p_{4}+p_{5})+(p_{1}-p_{5})
-(p_{1}+\sqrt{2}\Pi_{P})
\nonumber
\\
p_{4}&=&p_{1}+(p_{3}+p_{4}+p_{5})+(p_{2}-p_{4})
-(p_{1}+\sqrt{2}\Pi_{P})
\label{29}
\end{eqnarray}
If measurement has outcome  $p, P, Q$ then in (\ref{28}) and (\ref{29})  it results in
$p_{1}\to p$, $\Pi_{P}\to P$ and $Y_{Q}\to Q$. Taking into account properties of
the teleported state given by (\ref{24}) and  GHZ (\ref{27}) one can find
the operators of positions of teleported state and GHZ to be connected
by unitary transformation
\begin{eqnarray}
x_{4}&=&x_{2}-\sqrt{2}Y_{Q}
\nonumber
\\
x_{5}&=&x_{1}-\sqrt{2}Y_{Q}
\label{30}
\end{eqnarray}
For the operators of total momentum it follows
\begin{equation}
p_{4}+p_{5}=p_{1}+p_{2}
\label{31}
\end{equation}
Thus entanglement of the  light modes 1 and 2 is reproduced by pair modes 4 and 5 of GHZ.

This work was supported in part by Delzell Foundation.

\end{document}